\documentclass[10pt,twocolumn,prl,aps,amssymb,amsmath,tightenlines,longbibliography]{revtex4-2} 

\usepackage[hidelinks]{hyperref}
\usepackage{dsfont}
\usepackage{graphicx}
\usepackage{amssymb,amsfonts,amsthm}
\usepackage{color}
\usepackage{bm,float}

\newcommand{\rmi}{{\rm i }}
\newcommand{\rme}{{ \rm e }}

\begin{document}

\title{Imaginary Barrier, Real Transfer: Non-Hermitian Dynamical Tunnelling}

\author {Eva-Maria Graefe} 
\affiliation{Imperial College London, London SW7 2BZ, UK }

\begin{abstract}
We identify a non-Hermitian counterpart of dynamical tunnelling induced by a localised absorbing region in a symmetric confining potential. The absorbing region, described by an imaginary potential barrier, induces an effective double-well structure with a pair of long-lived states that have little population in the barrier. In contrast to the periodic tunnel oscillations through a real barrier, the system slowly evolves towards equal population of the two sides, while the absorbing region remains only weakly populated. We demonstrate this generic mechanism in two continuous potentials, and capture its essence analytically in a minimal $3\times 3$ matrix model. We propose a realistic optical waveguide setup for the experimental observation of the effect.
\end{abstract}

\maketitle

In Hermitian quantum mechanics, a potential barrier embedded in a symmetric confining potential leads to the hallmark phenomenon of tunnelling \cite{raza2013book}. A symmetric double-well potential gives rise to pairs of symmetric and antisymmetric states localised away from the barrier, with near degenerate energies. An initial superposition of these doublet states exhibits a characteristic slow periodic transfer between the two wells, with only minimal population in the barrier. The concept of dynamical tunnelling associated to a symmetric/antisymmetric pair of states with near degenerate energies also underlies more general processes with transfer between distinct, dynamically separated, classical phase space regions \cite{Davi1981,kesh2011book}.

A characteristic feature of non-Hermitian quantum systems with localised loss is the restructuring of states into short-lived states localised in the loss regions, and long-lived states away from it \cite{pers2000,okol2003,rott2009,guo2009,dave2024,chai2026}. In this sense, localised loss can act as an effective barrier and a strong imaginary potential can act similarly to a real one \cite{rama2003,burk2020}. 
Given that a real barrier in a symmetric confinement gives rise to tunnelling, this raises a natural question: What is the counterpart of tunnelling when the barrier is imaginary rather than real? 

Here we show that, while there is no real barrier in place, perhaps surprisingly, a particle initially localised on one side of the trapping potential ``tunnels'' partially to the other side without significant occupation in the middle lossy region—this is what we will refer to as non-Hermitian tunnelling. Just as in the Hermitian case, this process is brought about by the occurrence of a symmetric/antisymmetric pair of states with near degenerate energies, the superposition of which is localised on one side of the barrier. In contrast to the periodic oscillations between the two sides of a real potential barrier in a confining potential, the superposition of the doublet states in the non-Hermitian case relaxes into the longer-lived state in the doublet, asymptotically approaching equal occupation of the two sides. 

We demonstrate this phenomenon for a quantum particle in a harmonic potential with an imaginary Gaussian barrier, and in an infinite square well potential with a central region of constant imaginary potential. We show that the essence of the process is captured by a minimal $3\times 3$ matrix model. Finally, we propose an experimental implementation inspired by the optical multi-mode waveguide setup in \cite{dave2024,chai2026}.

\begin{figure*}[htb]
      \centering
 \includegraphics[width=0.95\textwidth]{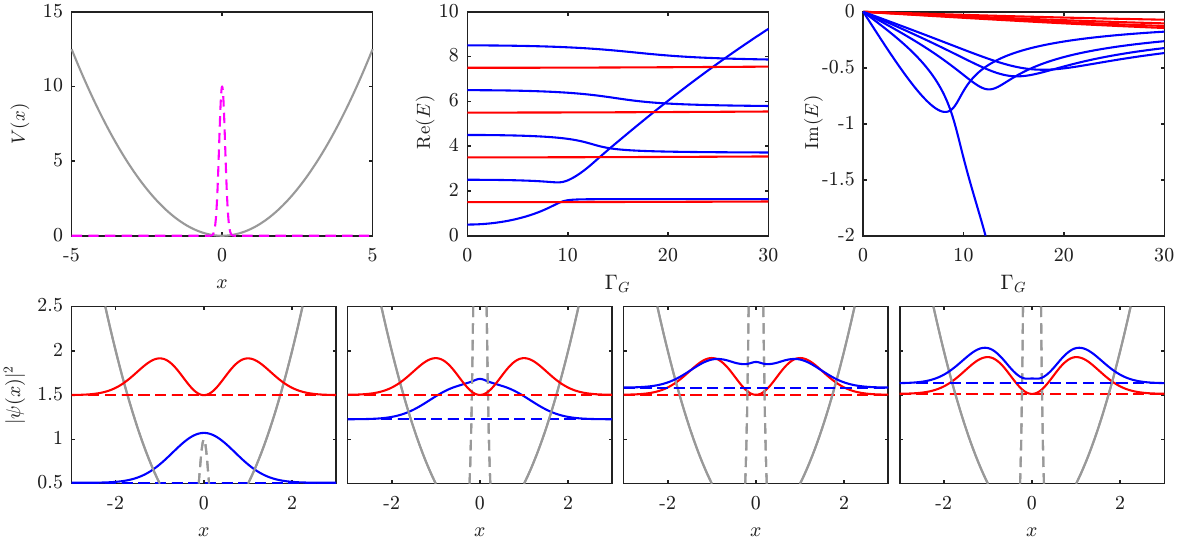}  
\caption{Harmonic trap with a central Gaussian imaginary barrier. 
Top row: left: Real confining potential $V(x)$ (solid grey) and imaginary barrier $\Gamma(x)$ (dashed magenta), for $\Gamma_{\! G}=10$ and $\sigma=0.1$; middle and right: real parts $\text{Re}(E)$ (middle) and imaginary parts $\text{Im}(E)$ (right) of the nine complex eigenvalues with the lowest real parts as functions of the peak absorption strength $\Gamma_{\! G}$, symmetric modes are plotted in blue, antisymmetric ones in red. Bottom row: Probability distributions in the two eigenstates with the lowest real part of the energy (even in blue, odd in red) depicted within the potential (dashed grey) for different values of $\Gamma_{\! G}$, from left to right: $\Gamma_{\! G}=1,\, 8,\, 10,\, 20$. }
\label{fig_HO_Gaussian_abs_barrier1} 
\end{figure*}

We begin by considering a quantum particle in a harmonic trap with a central Gaussian-shaped absorption profile, modelled by the Hamiltonian
\begin{equation}
\hat K=\frac{1}{2}\hat p^2+\frac{1}{2}\hat q^2-\rmi\Gamma_{\!G}\,\rme^{-\frac{1}{2\sigma^2}\hat q^2}.
\end{equation}
We use natural units with $\hbar=m=\omega=1$ (where $\omega$ is the frequency of the harmonic confinement). Thus, energy is measured in units of $\hbar\omega$ and all variables are dimensionless numerical values.  The peak absorption strength $\Gamma_{\!G}$ is assumed real and positive, and $\sigma^2$ is the dimensionless width parameter of the absorption profile. The real and (negative) imaginary parts of the potential are depicted for $\sigma=0.1$ and 
$\Gamma_{\!G}=10$ in the top left panel of Fig.\ref{fig_HO_Gaussian_abs_barrier1}. The other two figures on the top show the real (middle) and imaginary (right) parts of the nine eigenvalues with the lowest real parts as functions of the peak absorption strength $\Gamma_{\! G}$. The eigenvalues belonging to even eigenfunctions are depicted in blue, those belonging to odd eigenfunctions in red. The bottom row of the same figure depicts the probability distribution for the two eigenfunctions with the lowest real parts of the energy for four increasing values of $\Gamma_{\! G}$. 

While the exact spectral behaviour depends on the value of $\sigma$, some features are generic. In particular, as $\Gamma_{\! G}$ is increased, the real parts of the eigenvalues stay approximately constant for small values of $\Gamma_{\! G}$, the states acquire negative imaginary parts, whose magnitudes initially increase linearly, as predicted by first order perturbation theory.  Due to the overlap with the absorption profile, the even eigenstates pick up a larger decay rate than the odd ones, and the ground state is most sensitive to small absorption. As the value of $\Gamma_{\! G}$ increases further a typical restructuring of the eigenstates often referred to as resonance trapping, takes place \cite{pers2000,okol2003,rott2009}.  Most eigenstates become long-lived, while a few (only the state corresponding to the second excited state in the Hermitian limit, amongst the energies depicted here) acquire a larger decay rate. For large values of the peak absorption, the absorbing profile acts as an effective quantum barrier that decouples the two sides of the harmonic trap; while some states localise in the absorbing region and become very short lived, the other lower energy states localise away from the absorbing region and form symmetric/antisymmetric doublets with near-degenerate complex energies. 

Thus, for relatively large values of $\Gamma_{\! G}$ an equal superposition of the symmetric and antisymmetric doublet states is localised primarily in one of the two sides of the potential well, away from the imaginary barrier. Dynamically, such a superposition performs a non-Hermitian generalisation of dynamical tunnelling composed of two competing processes; the typical Hermitian periodic oscillations between the symmetric and antisymmetric state, resulting in a periodic oscillation between the two sides of the imaginary barrier, and a slow relaxation into the longer-lived state, where the population is equally distributed between the two halves of the potential. Throughout the dynamics the population within the bulk of the absorbing barrier remains small. The timescales for each of these processes are dictated by the real and imaginary parts of the energy splitting, i.e. the oscillation frequency is given by the real part of the energy splitting, and the relaxation rate into the more stable state is given by the imaginary part of the splitting. Because the absorbing barrier has a finite tail, no eigenstate has infinite lifetime, and the overall probability decays over time. 

\begin{figure}[htb]
      \centering
 \includegraphics[width=0.49\textwidth]{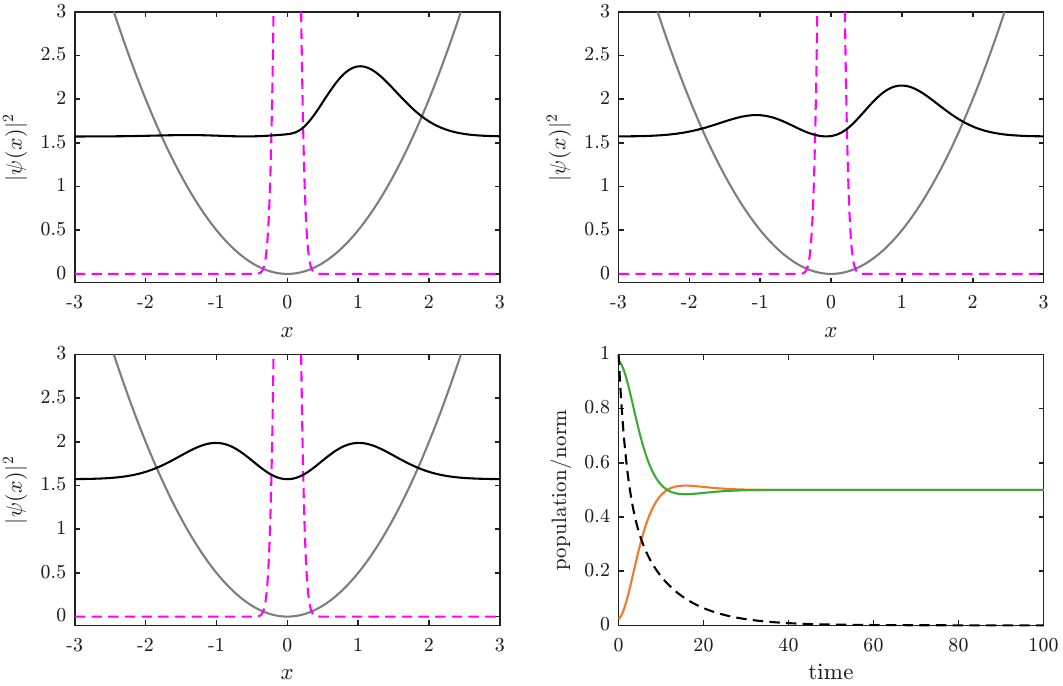}  
\caption{Non-Hermitian tunnelling in a harmonic trap with a central Gaussian imaginary barrier.  Snapshots of the (normalised) probability distribution plotted at the height of the expected value of the real energy in the external potential for $\sigma=0.1$ and $\Gamma_{\! G}=20$ for three different times ($t=0$ (top left), $t=5$ (top right), and $t=100$ (bottom left)); bottom right: time evolution of the probability in the right (green) and left (orange) half of the potential and total norm (dashed black).
}
\label{fig_HO_Gaussian_abs_barrier_dyn} 
\end{figure}

The non-Hermitian effect of relaxation into equal population of the two sides away from the imaginary barrier is most obvious when the gap in imaginary parts is larger than that in the real parts. In Fig. \ref{fig_HO_Gaussian_abs_barrier_dyn} we illustrate the effect for $\sigma=0.1$ and $\Gamma_{\! G}=20$ for an initial state that is an equal superposition of the doublet states, numerically propagated using a split operator method. 
The top row and the plot on the bottom left show snapshots of the probability distribution at three different times ($t=0,\,5,$ and $100$). The right plot in the bottom row shows the time evolution of the probability of detection in half of the potential well (green line: right half, orange line: left half), as well as the overall norm (dashed black line). We clearly observe the relaxation into the final state with equal probability on each side. Here the most stable state is the antisymmetric state, which is the analytical continuation of the first excited state. Due to the difference in the real parts of the doublet energies, there is a small remnant of real tunnel oscillations that is visible in the small dip/peak in the population of the two sides for intermediate times around $t\approx 10$. We further observe the expected decay of the overall probability.

This non-Hermitian tunnelling behaviour is in fact typical for symmetric confining potentials with imaginary barriers. To illustrate this, let us consider a second model, given by a quantum particle in an infinite square well potential with a central absorbing region, described by the 
Hamiltonian
\begin{equation}
\hat K=\frac{1}{2m}\hat p^2+V(\hat q),
\end{equation}
with the complex potential
\begin{equation}
V(x)=\left\{ \begin{array}{ll} 
\infty,\quad x \leq -L \\ 
0,\quad -L<x\leq -a,\\
-\rmi\Gamma_0,\quad -a< x< a,\\
0,\quad a\leq x< L,\\
\infty,\quad L \leq x,
\end{array} \right. 
\end{equation}
where $0<a<L$, and $\Gamma_0$ is a positive real absorption constant.

Just as in the Hermitian textbook case, 
from the boundary conditions $ \phi(\pm L)=0$ and the continuity of the wavefunction 
and its first derivative at $x=\pm a$, one obtains the quantisation conditions
\begin{equation}
k \cos((L-a)k)\cos(a\tilde k)=\tilde k\sin(a\tilde k )\sin((L-a)k)
\end{equation}
 and 
\begin{equation}
k \cos((L-a)k)\sin(a\tilde k)=-\tilde k\cos(a\tilde k )\sin((L-a)k),
\end{equation}
for even and odd states, respectively, where $k=\frac{\sqrt{2mE}}{\hbar}$ and $\tilde k=\frac{\sqrt{2m(E+\rmi\Gamma_0)}}{\hbar}$.

\begin{figure}[htb]
      \centering
 \includegraphics[width=0.49\textwidth]{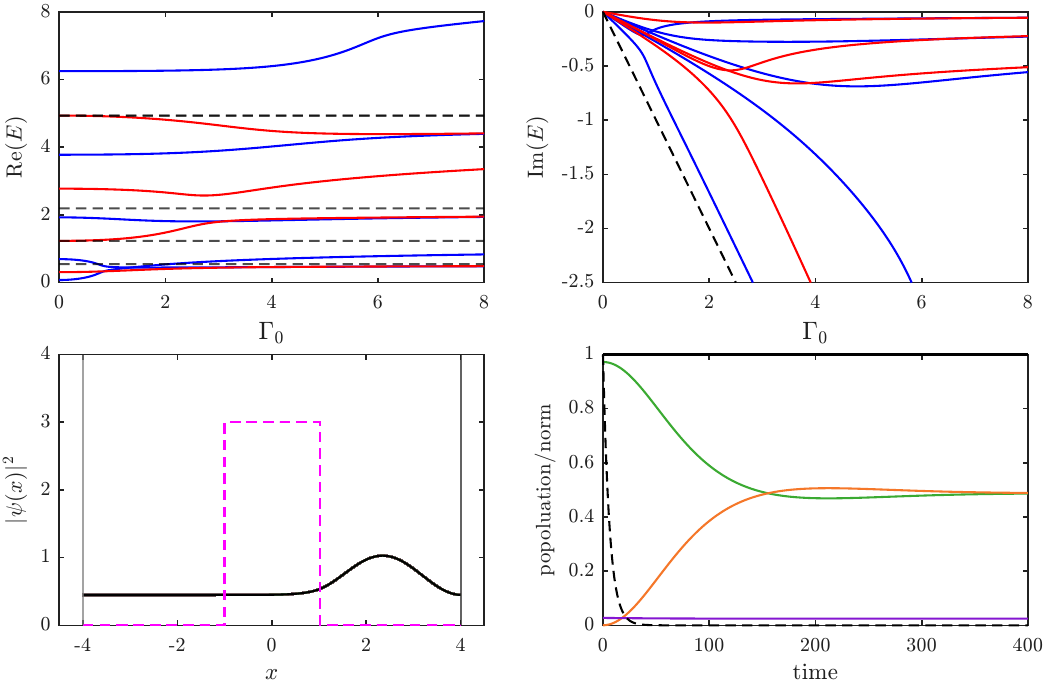}  
\caption{Non-Hermitian tunnelling in a box of size $2L=8$ with a central rectangular imaginary barrier of width $2a=2$. Top: Real (left) and imaginary (right) parts of eigenvalues as functions of $\Gamma_0$,  dashed black lines indicate the asymptotic eigenvalues for large $\Gamma_0$; bottom left: Equal superposition of the doublet states, localised in the right half of the effective double well for $\Gamma_0=3$. The box boundaries are indicated by grey lines, and the absorption profile by a dashed magenta line; bottom right: time evolution of the probability in the right (green) and left (orange) wells, and in the absorbing region (purple), and total norm (dashed black).
}
\label{fig_box_imag_region} 
\end{figure}

With increasing height of the absorbing barrier, each of the solutions of the Hermitian well deforms continuously into a new solution of the non-Hermitian problem associated with a complex value of $k$ and hence a complex energy. 
In the limit $\Gamma_0\to\infty$ the three regions decouple; both quantisation conditions reduce to either $\sin((L-a)k)=0$, with solutions given 
by the eigenstates of two separate potential wells of length $L-a$ with hard walls, with doubly degenerate real eigenenergies 
\begin{equation}
E_{j,\text{ real}}^{(\infty)}=\frac{\hbar^2\pi^2 j^2}{m(L-a)^2}
\end{equation}
or $\sin(a\tilde k)=0$ or $\cos(a\tilde k)=0$, that is $\tilde k_n=\frac{n\pi}{a}$, corresponding to eigenenergies 
\begin{equation}
E_{j,\text{ decay}}^{(\infty)}=\frac{\hbar^2\pi^2 j^2}{8ma^2}-\rmi\Gamma_0,
\end{equation}
 i.e. very short-lived states, identical to the eigenstates of a real box of size $2a$.
  
For intermediate values of $\Gamma_0$ the eigenvalues and eigenstates can be obtained from a numerical solution of the quantisation conditions. As an example, the eigenvalues with the nine lowest real parts are depicted as functions of $\Gamma_0$ in the top panels of Fig.~\ref{fig_box_imag_region} for $L=4$ and $a=1$. We observe a qualitatively similar behaviour to the harmonic oscillator with the Gaussian absorbing barrier, with the typical restructuring of the spectrum with increasing absorption rate $\Gamma_0$ and the formation of the non-Hermitian doublets. The bottom panel on the left shows the probability distribution of an equal superposition of the lowest energy doublet for $\Gamma_0=3$, localised mainly in the right effective well. The time-dependent populations in each region of the well, and the overall norm for this initial state are depicted in the right bottom panel. The numerical propagation was performed using a finite-difference simulation. We observe the characteristic slow relaxation into equal population of the outer two wells with only a small near-constant population in the central absorbing region. In contrast to the first example, here the asymptotic state is symmetric. The quantitative features of the dynamics and the parity of the asymptotic state depend on the exact system parameters. 

\begin{figure}[t!]
      \centering
\includegraphics[width=0.48\textwidth]{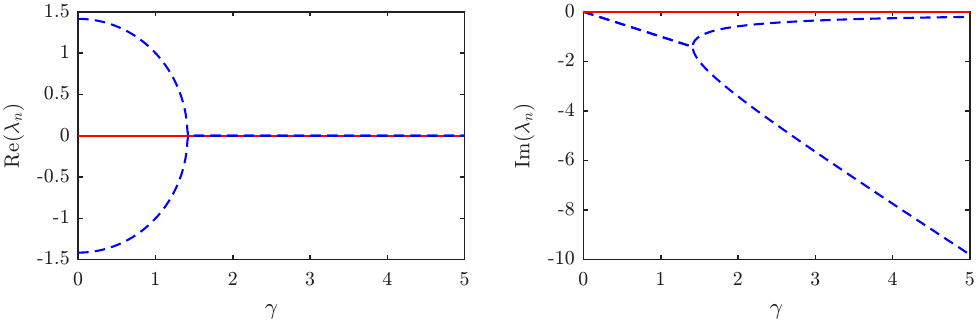} 
\caption{Real (left) and imaginary (right) parts of the eigenvalues of the minimal matrix model as functions of the loss parameter $\gamma$, for unit coupling $v=1$. The energies of the two symmetric states are displayed as dashed blue lines, that of the antisymmetric state as solid red line. }
\label{fig_3x3_eig} 
\end{figure}

The essence of this non-Hermitian tunnelling phenomenon can be captured by 
a simple matrix model with three states, representing the left, central (absorbing), and right regions, described by a Hamiltonian 
of the form
\begin{equation}
\label{eqn-Ham3x3}
\hat K=\left(\begin{array}{ccc}
0 & -v & 0 \\
-v & -2\rmi\gamma & -v  \\
0 & -v & 0
\end{array}\right),
\end{equation}
where $v$ and $\gamma$ are real and positive constants. Such matrix Hamiltonians also arise as effective models of three coupled waveguides with absorption in the central guide, that have been studied in the contexts of anti-PT symmetry and passive beam splitting \cite{yang2017,fan2020,alri2021}. The eigenvalues are $\lambda_0=0$ and $\lambda_\pm=-\rmi\gamma\pm\sqrt{2v^2 - \gamma^2}$, with corresponding eigenvectors
\begin{equation}
\label{eqn_3x3eigvec}
\phi_0=\begin{pmatrix}1\\0\\-1\end{pmatrix},\quad \phi_\pm=\begin{pmatrix}                                
 v\\ \lambda_\pm\\  v\end{pmatrix}.
\end{equation}
The real and imaginary parts of the eigenvalues are depicted as a function of the decay rate $\gamma$ for $v=1$ in Fig.~\ref{fig_3x3_eig}. The eigenvector $\phi_0$ has no population in the central site, and the corresponding eigenvalue stays zero independently of the value of $\gamma$. The other two eigenvectors and eigenvalues display the typical resonance trapping behaviour, associated to an exceptional point at $\gamma=\sqrt{2}v$, where the eigenvectors $\phi_\pm$ coalesce. 

For $\gamma\gg v$, the eigenvalues are approximately given by
\begin{equation}
\lambda_0=0,\quad \lambda_+\approx -\rmi\frac{v^2}{\gamma},\quad \lambda_-\approx -2\rmi\gamma.
\end{equation}
Introducing the small parameter  $\epsilon=v/\gamma\ll 1$ 
the eigenvectors (normalised in leading order of $\epsilon$) are approximately given by
\begin{equation}
\label{eqn_eigvec_3x3_gam_large}
\phi_0=\frac{1}{\sqrt{2}}\begin{pmatrix}1\\0\\-1\end{pmatrix},\quad\phi_+ \approx\frac{1}{\sqrt{2}}\begin{pmatrix}1\\ -\rmi\epsilon\\1\end{pmatrix},\quad  \phi_-\approx\begin{pmatrix}\epsilon\\ \rmi\\ \epsilon\end{pmatrix}.
\end{equation}
An initial state $\psi_0=(1,0,0)^T$, localised entirely in the left region, is expanded into the eigenbasis (\ref{eqn_eigvec_3x3_gam_large}) as
 \begin{equation}
\psi(0)\approx\frac{1}{\sqrt{2}}\left(\phi_0+\phi_+ + \frac{\epsilon}{\sqrt{2}}\phi_-\right),
\end{equation}
and thus the  time-evolution is given by 
\begin{align}
\nonumber \psi(t)&\approx\frac{1}{\sqrt{2}}\left(\phi_0+\rme^{-\gamma\epsilon^2 t}\phi_++\frac{\epsilon}{\sqrt{2}}\rme^{-2\gamma t}\phi_-\right)\\
&\approx\frac{1}{2}\begin{pmatrix}1+\rme^{-\gamma\epsilon^2 t}\\ \frac{\epsilon}{\sqrt{2}}\rme^{-2\gamma t}\\ -1+\rme^{-\gamma\epsilon^2 t}\end{pmatrix}.
\end{align}
That is, the initial state $(1,0,0)^T$ continuously (and slowly) evolves into the antisymmetric state $(1,0,-1)^T$, while losing half of its overall norm. This is the essence of the non-Hermitian tunnelling behaviour introduced here.

Let us conclude with proposing a realistic experimental setup for the observation of the effect utilising multimode waveguides with metal cladding \cite{dave2024,chai2026}. Modelling the cross section of such a waveguide by an effective 1-dimensional refractive index profile, and employing a paraxial approximation, the propagation of light is described by a Schr\"odinger-like equation of the form \cite{long2009}
\begin{equation}
\rmi\lambdabar\frac{\partial}{\partial z}\psi(x,z)\!=\!\left(\!-\frac{\lambdabar^2}{2n_s}\frac{\partial^2}{\partial x^2}-\frac{n_s^2-n(x)^2}{2n_s}\right)\!\psi(x,z).
\end{equation}
Here $z$ denotes the direction of the waveguide and the main propagation direction, playing the role of time in the Schr\"odinger equation. The substrate refractive index is denoted by $n_s$, and the refractive index profile $n(x)$ only varies in $x$ direction. The envelope of the electric field amplitude for the transverse electric mode $\psi(x,z)$ takes the role of the wave function, and we have $\lambdabar=\lambda/2\pi$, where $\lambda$ is the vacuum wavelength of the electromagnetic wave. The variation of the refractive index acts as an analogue of a potential. 

\begin{figure*}[htb]
      \centering
 \includegraphics[width=0.95\textwidth]{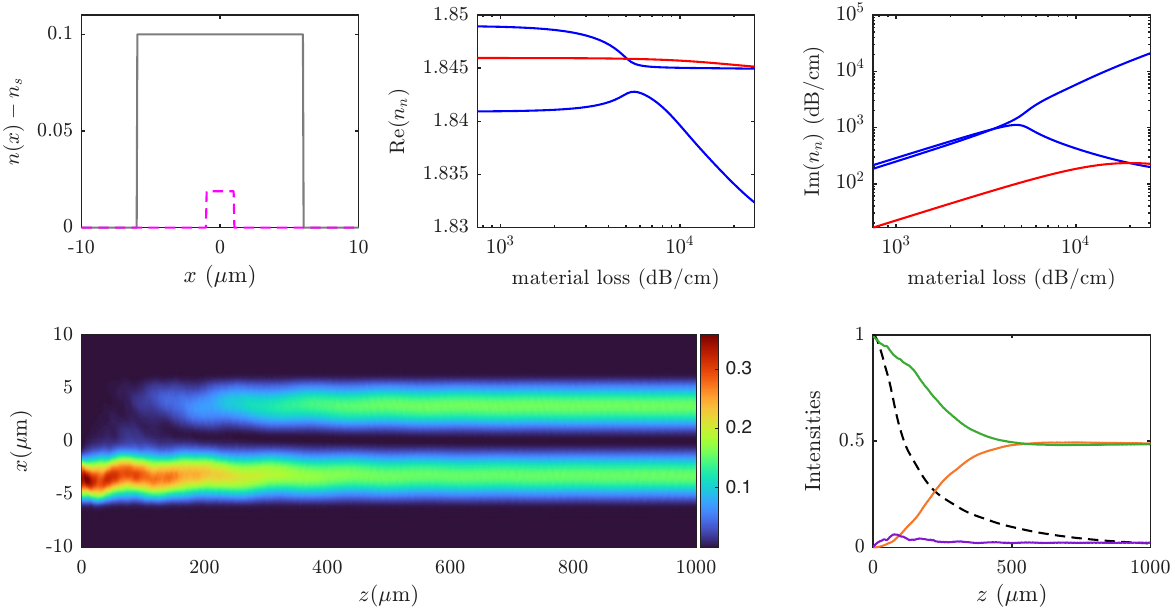} 
  \caption{Model of multimode waveguide with central absorbing region. Top row: left: Real (solid grey) and imaginary part (dashed magenta) of the refractive index profile; middle and right: real parts (middle) and imaginary parts (right) of the TE mode refractive indices as functions of absorption for $\lambda=1.55\mu{\rm m}$, for the three modes connected to the three lowest TE modes in the non-absorbing case. Symmetric modes are plotted in blue, antisymmetric ones in red. Bottom row: Propagation of an initial mode corresponding to the fundamental mode of a waveguide in the region $[-6, -1]$ for ${\rm Im}(n)=-1.78\times10^{-2}\approx 7\times 10^3\,\text{dB/cm}$ in the absorbing region. Left panel: False colour plot of the normalised square amplitude of the electromagnetic field; right panel:  population in the right (green) and left (orange) effective wells, and in the absorbing central region (purple), and total norm (dashed black) as functions of the propagation distance.}
\label{fig_waveguide} 
\end{figure*}

Specifically, we consider the refractive index profile sketched in the top left in Fig.~\ref{fig_waveguide} with background refractive index of $n_s=1.75$, corresponding to an $\mathrm{LiNbO}_3$ base, and a single waveguide of width $12\mu{\rm m}$ with $n_{wg}=1.85$, corresponding to a silicon nitride (SiN) cladding. An absorbing region of width $2\mu{\rm m}$ with imaginary refractive index of ${\rm Im}(n)\approx-10^{-2}$ can be implemented by an additional strip of metal on top of the central part of the SiN cladding \cite{dave2024,chai2026}.

The real and imaginary parts of the effective refractive indices of the three modes arising as a continuation of the three lowest energy modes in the absence of the absorbing region are plotted as functions of the imaginary part of the refractive index (i.e. the absorption strength) in the central waveguide in the middle and right top panels, for a wavelength $\lambda=1.55\mu{\rm m}$. In the bottom row we depict the propagation of the (renormalised) electric field amplitude for ${\rm Im}(n)=-1.78\times10^{-2}\approx 7\times 10^3\,\text{dB/cm}$ in the absorbing region, for an incoupled mode corresponding to the eigenmode of a single waveguide between $x=-6$ and $x=-1$ (the left uncladded part of our setup), corresponding to the experimental setup in \cite{dave2024,chai2026}. The left plot shows a false colour plot of the spatial intensity, the right plot shows the total intensity (black dashed), as well as the intensity in the three waveguide regions (green, orange, and purple, respectively). The initial mode profile contains not only the non-Hermitian tunnelling doublet, but also some contributions of higher modes, leading to small spatial oscillations in the $x$ direction in the intensity profile. Nevertheless, we clearly observe the characteristic non-Hermitian tunnelling behaviour on experimentally realistic length scales of $100\mu{\rm m}--1{\rm cm}$, with only moderate overall loss.

In summary, we have identified a non-Hermitian counterpart of dynamical tunnelling induced by a localised absorbing barrier in a symmetric confinement. The mechanism is generically found in various continuous models and is captured by a minimal $3\times 3$ matrix model. We have proposed a realistic experimental setup for the observation of the effect. 

\begin{acknowledgments}
The author acknowledges financial support from the EPSRC through the program grant ``Next generation metamaterials: exploiting four dimensions'' (Grant No.~EP/Y015673/1). Stimulating discussions with Dorje C. Brody and Jessica Eastman are gratefully acknowledged.
\end{acknowledgments}

\bibliography{bibliography}

\end{document}